\documentclass[twocolumn,10pt,aps,prd]{revtex4-1}

\usepackage{amssymb,amsmath,amsthm}
\usepackage[pdftex]{graphicx} 
\usepackage{braket}
\usepackage[T1]{fontenc}
\usepackage[latin1]{inputenc}
\usepackage[english]{babel}
\usepackage[sort&compress]{natbib}

\usepackage{float}
\usepackage{braket}
\usepackage{slashed}
\usepackage{commath}
\usepackage{diagbox}
\usepackage{color}
\usepackage{mathrsfs}
\usepackage{bm}
\usepackage{hyperref} 
\hypersetup{colorlinks=true,citecolor=blue,linkcolor=blue,urlcolor=blue}
\usepackage{booktabs}

\usepackage{tikz}
\usetikzlibrary{positioning,arrows,patterns}
\usetikzlibrary{decorations.markings} 
\usetikzlibrary{decorations.pathmorphing}

\addto\captionsenglish{}

\addto\captionsenglish{}

\renewcommand{\epsilon}{\varepsilon}
\newcommand{\mb}[0]{\mathbf}

\renewcommand{\rm}[1]{\mathrm{#1}}

\newcommand{\p}{\partial}
\newcommand{\w}{w^{\rm{add}}}
\newcommand{\wt}{w^{\rm{tot}}_{\phi\to\psi\psi}}

\newcommand{\J}[2]{J_{#1}(#2)}
\newcommand{\Y}[2]{Y_{#1}(#2)}

\begin{document}
\title[]{Decaying Massive Particle in Matter and Radiation Dominated Eras}

\author{Juho Lankinen}
\email{jumila@utu.fi}
\affiliation{Turku Center for Quantum Physics, Department of Physics and Astronomy, University of Turku, 20014 Turku, Finland}

\author{Iiro Vilja}
\email{vilja@utu.fi}
\affiliation{Turku Center for Quantum Physics, Department of Physics and Astronomy, University of Turku, 20014 Turku, Finland}

\begin{abstract}
According to the standard model of cosmology, the early universe has been dominated by radiation or non-relativistic matter in several eras of its history. However, many cosmological calculations involving particle processes are commonly done using Minkowskian results for them, although, for more precise treatment, quantum field theory in curved spacetime is needed. This paper aims to fill this gap by presenting decay rates for matter and radiation dominated universes in this more precise treatment. We provide a study of the average decay rates for a process where a conformally coupled massive scalar field decays into massless scalar particles. It is found that the presence of a curved spacetime modifies the Minkowskian result considerably for early times but asymptotically only by an additive term proportional to the inverse of mass and interaction time. Thus, the correction is small for large time scales, but on the time scales of the order of $m\sim t$, the relative correction term may be of importance.
\end{abstract}

\maketitle

\section{Introduction}
Cosmological calculations involving decay rates have commonly been done by using the decay rate obtained in flat space. The reason behind this has mainly been the fact that calculating the decay rate in curved spacetime is highly non-trivial. Therefore, the question of how good this Minkowskian approximation really is is open to questions since, for a precise treatment, the calculations for the decay rate have to be done using the machinery of quantum field theory in curved spacetime.
In this theory one is confronted with gravitational particle creation, found some decades ago \citep{Parker:1968,Parker:1969,Zeldovich_Starobinsky:1972,Zeldovich_Starobinsky:1977}, which interferes with the decay process making the calculation of the decay rate much more complicated. Although self-interacting field theories have been investigated in great detail, the theory behind mutually interacting fields and decay rate calculations still remains far from complete.
Few works devoted to this were done some time ago \citep{Audretsch_Spangehl:1985,Audretsch_Spangehl:1986,Audretsch_Spangehl:1987,Birrell_Davies_Ford:1980}, but recently interest in mutually interacting fields in curved space has resurfaced \citep{Crucean_Baloi:2015,Blaga:2015,Lankinen_Vilja:2017b} along with some new ideas for dealing with the decay rate calculation \citep{Boyanovsky:2011,Boyanovsky:2012}. Aside from theoretical aspects, practical calculations for use in relevant cosmological situations have gone largely unnoticed.

According to standard cosmology, the early universe was an extremely hot place and in cooling down the Universe would have gone through several phase transitions.  
Indeed, in the first moments of the evolution of our universe, inflation may have occured to explain some of the problems faced with standard cosmology \citep{Guth:1981,Linde:1982}. Inflation itself may have ended in yet another phase, known as kination or deflation period, for a short time \citep{Peebles_Vilenkin:1999,Spokoiny:1993,Joyce:1997}. There is, however, no experimental verification for these eras. What is known though, is that the Universe was once dominated by radiation until it gave way for a ordinary matter dominated era, making them the most relevant phases when considering cosmological situations.

In this article we study particle decay in matter and radiation dominated eras. Specifically, we confine our investigation to an interaction where a massive scalar particle decays into two massless scalar particles both of which are conformally coupled to gravity. This prescription allows us to use the concept of added-up probability and calculate the transition probability of this process exactly. First, we derive a general formula for the total transition probability for a given flat Friedmann-Robertson-Walker-metric scale factor $a$ and given massive field mode $\chi$ in its rest frame. Secondly, we study particle decays properly. It is found that, asymptotically, the Minkowskian decay rates are modified by the expansion of spacetime by an additive correction which depends on the inverse of mass. Moreover, the decay of a scalar particle is slower compared to Minkowski space. We compare these results with the previously calculated decay rate for a universe filled with stiff matter \citep{Lankinen_Vilja:2017b}. The correction term in all three cases is small implying that, at least on large time scales, the Minkowskian decay rates give a good enough approximation in calculations of decay rates on these spacetimes at least for the particular process in question.

This paper is organized as follows. In Sec. \ref{sec:II} we give an introduction into the formalism needed in calculation of the decay rate and derive a general transition probability equation. Sections \ref{sec:III} and \ref{sec:IV} contain the calculations of the decay rates for radiation and matter dominated eras, respectfully. Section \ref{sec:V} contains discussion. Throughout this paper we work in natural units $\hbar=c=1$ and the metric is chosen with positive time component.

\section{Preliminaries}\label{sec:II}
\subsection{Theoretical framework}
The four-dimensional spatially flat Friedmann-Robertson-Walker spacetime is described by the metric
\begin{align}
ds^2=a(\eta)^2(d\eta^2-d\bm{\mathrm{x}}^2)
\end{align}
given in conformal time $\eta\in (0,\infty)$, where the expansion of spacetime is characterized by the dimensionless scale factor $a(\eta)$. We consider a massive real scalar field $\phi$ and a massless real scalar field $\psi$ which are conformally coupled to gravity. The Klein-Gordon equation in this case is
\begin{align}\label{Klein-Gordon}
(\square+m^2+R/6)\phi(\eta,\mb x)=0,
\end{align}
where $\square$ denotes the covariant d'Alembert operator and $R$ is the Ricci scalar. For massless fields take $m=0$. Because of the homogeneity of the spatial sections, the mode solutions $u_\mb k$ of Eq. \eqref{Klein-Gordon} are separable,
\begin{align}\label{Massive_mode_solutions}
u_\mb p(\eta,\mb x)=\frac{e^{i\mb p\cdot \mb x}}{(2\pi)^{3/2}a(\eta)}\chi_p(\eta),
\end{align}
where $p:=|\mb p|$. For a conformally coupled field $\chi_p$ satisfies the equation
\begin{align}\label{KG_chi}
\chi_p''(\eta)+(p^2+a(\eta)^2m^2)\chi_p(\eta)=0.
\end{align}
The mode solutions can be obtained by solving Eq. \eqref{KG_chi} and using the asymptotic condition to recognize the positive modes in the standard way \cite{birrell_davies_1982}. For a massless field, the corresponding mode solutions are obtained straightforwardly from the flat space solutions,
\begin{align}\label{Massless_mode_solution}
v_\mb{k}(\eta, \mb x)&=\frac{1}{(2\pi)^{3/2}a(\eta)}\frac{e^{i\mb{k\cdot x}-ik\eta}}{\sqrt{2k}},
\end{align}
where $k:=|\mb k|=k^0$.
The interaction between the massive and massless fields is given by the Lagrangian
\begin{align}\nonumber
\mathcal{L}=&\frac{\sqrt{-g}}{2}\big\{\p_\mu \phi \p^\mu\phi-m^2\phi^2-\frac{R}{6}\phi^2+\p_\mu\psi\p^\mu \psi-\frac{R}{6}\psi^2\big\}\\
&+\mathcal{L_I},
\end{align}
where $g$ stands for the determinant of the metric. For the interaction term, we choose
\begin{align}\label{L_I}
\mathcal{L}_I=-\sqrt{-g}\lambda \phi \psi^2,\ \lambda>0.
\end{align}
The $S$-matrix is given as
\begin{align}
S=\lim_{\alpha\to 0^+}\hat{T}\exp\Big(i\int\mathcal{L_I}e^{-\alpha\eta}d^4x\Big),
\end{align}
where $\hat{T}$ denotes the time-ordering operator. The exponential factor $e^{-\alpha\eta}$ acts as a switch off for the interaction for large times with $\alpha$ being a positive constant and called the switch-off parameter.
The perturbative expansion of the $S$-matrix for the interaction \eqref{L_I} gives
\begin{align}
S=1-i\lambda A+\mathcal{O}(\lambda^2)
\end{align}
with
\begin{align}\label{A_integraali}
A:=\lim_{\alpha\to 0^+}\int \hat{T}\phi \psi^2 e^{-\alpha\eta}\sqrt{-g}\, d^4x
\end{align}
and we consider only tree level processes for which the transition amplitude is defined as 
\begin{align}
\mathscr{A}:= \braket{\rm{out}|A|\rm{in}}.
\end{align}

\subsection{Added-up probability}
\begin{figure}[H]
\centering
\includegraphics[scale=1]{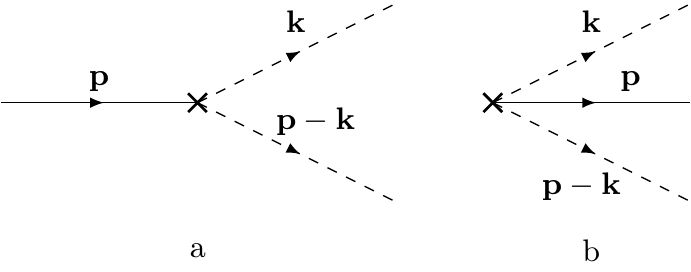}
\caption{Diagrams contributing to the added-up decay probability. The solid line corresponds to the massive particle and the dashed lines to massless particles.}\label{fig:1}
\end{figure}

The calculation of the decay rates and transition amplitudes in curved spacetime is a non-trivial task, since in curved spacetime the gravitational particle creation interferes with the process of mutual interaction. There exists, however, a couple of ways to calculate the decay rate \citep{Audretsch_Spangehl:1985,Boyanovsky:2011,Boyanovsky:2012}. In this article we will use the concept of added-up probabilities which requires that the massive particle decays into massless particles. Since conformally coupled massless particles are not created in an expanding conformally flat spacetime, if registered by a particle counter they have solely been created or influenced by the decay process. Furthermore, restricting only to those massive modes which fulfill the three-momentum conservation law $\mb p=\mb k_1+\mb k_2$, one is left with what resembles closest a decay process. This concept, known as added-up probability, was introduced by Audretsch and Spangehl in Ref. \cite{Audretsch_Spangehl:1985}, see also Ref. \citep{Lankinen_Vilja:2017b} for more details. 
In the added-up formalism,  the transition probability is given by
\begin{align}\nonumber\label{eq:w_add}
\w_{\phi\rightarrow \psi\psi}(\mb p,\mb k,\mb{p-k})=&\lambda^2\Big\{ \lvert \braket{\rm{out}, 1^\psi_{\mb{k}}1^\psi_{\mb{p-k}}\lvert A \rvert 1^\phi_{\mb{p}}, \rm{out}}\rvert^2\\
&+\lvert\braket{\rm{out},1^\phi_{\mb{-p}} 1^\psi_{\mb{k}}1^\psi_{\mb{p-k}}\lvert A \rvert 0, \rm{out}}\rvert^2 \Big\},
\end{align}
where $\mb k_1=\mb k$ and $\mb k_2=\mb{p-k}$. The corresponding Feynman diagrams are given in Fig. \ref{fig:1}.
\noindent The first term in \eqref{eq:w_add} corresponds to diagram a of Fig. \ref{fig:1}, while the second term refers to diagram b. 
The total decay probability is obtained by summing over all the $k$-modes,
\begin{align}\label{eq:w_tot}
w^{\rm{tot}}_{\phi\rightarrow \psi\psi}=\sum_{\mb k} \w_{\phi\rightarrow \psi\psi} (\mb p,\mb k,\mb{p-k}).
\end{align}
In what follows, we derive the equation for total transition probability for a general rest frame field mode with conformal coupling using the added-up formalism.

\subsection{Total transition probability}
Defining $\mathcal{A}^a(\mb k_1,\mb k_2,\mb p):=-i\lambda\braket{\rm{out},1_{\mb{k}_1}^\psi 1_{\mb{k}_2}^\psi|A|1_{\mb{p}}^\phi,\rm{out}}$ for the amplitude of diagram a and $\mathcal{A}^b(\mb k_1,\mb k_2,\mb q):=-i\lambda\braket{\rm{out},1^\phi_{\mb q}1_{\mb{k}_1}^\psi 1_{\mb{k}_2}^\psi|A|0,\rm{out}}$ for diagram b, the transition amplitude of diagram a of Fig. \ref{fig:1}, with the solutions \eqref{Massive_mode_solutions} and \eqref{Massless_mode_solution} is
\begin{align}\nonumber
\mathcal{A}^a(\mb k_1,\mb k_2,\mb p)&=\frac{-i\lambda \delta(\mb p-\mb k_1-\mb k_2)}{(2\pi)^{3/2} 2\sqrt{k_2k_2}}\\
&\times\lim_{\alpha\to 0^+}\int_0^\infty e^{-\alpha\eta}a(\eta)e^{i(k_1+k_2)\eta}\chi_p(\eta)d\eta.
\end{align}
For diagram b, the same amplitude is obtained with the changes $\delta(\mb p-\mb k_1-\mb k_2)\to\delta(\mb q+\mb k_1+\mb k_2)$ and $\chi_p(\eta)\to \chi_p(\eta)^*$. 
Performing the $\mb k_2$ integration and passing to continuum limit, we have for the total probability
\begin{align}
\wt=&\int_{\mathbb{R}^3} d^3\mb k\big(|\mathcal{A}^a(k,|\mb p-\mb k|)|^2+|\mathcal{A}^b(k,|\mb p-\mb k|)|^2\big).
\end{align}
Next, we go into the rest frame of the massive particle $\mb p=0$ and use spherical coordinates to obtain
\begin{align}\label{wtEq}
\wt=\frac{\lambda^2}{8\pi^2}\int_{-\infty}^\infty dk \Big| \int_0^T a(\eta)e^{2ik\eta}\chi_{p=0}(\eta) d\eta \Big|^2.
\end{align}
To obtain this equation, we have used the fact that $|\mathcal{A}^a(k,|\mb p-\mb k|)|^2=|\mathcal{A}^b(-k,-|\mb p-\mb k|)|^2$. We have also introduced a cutoff at $\eta=T$ and taken the $\alpha$ limit inside the integral.
Equation \eqref{wtEq} is essentially a three-dimensional integral containing the $k$ integration and a two-dimensional integration from the absolute value. The $k$ integral can be treated as a distribution, since it contains an integral of the form
\begin{align}\label{eq:delta_function}
\int_{-\infty}^\infty e^{i k(x-y)}dk=2\pi\delta(x-y).
\end{align}
Using this, the total transition probability can be cast in the form
\begin{align}\nonumber
\wt=&\frac{\lambda^2}{8\pi}\int_0^T\int_0^Td\eta d\eta'\delta(\eta-\eta')a(\eta)a(\eta')\\
&\times\chi_{p=0}(\eta)\chi_{p=0}(\eta')^*.
\end{align}
Performing the delta integration we obtain a general form for the total transition probability,
\begin{align}\label{Result1}
\wt=\frac{\lambda^2}{8\pi}\int_0^T a(\eta)^2\big|\chi_{p=0}(\eta)\big|^2 d\eta.
\end{align}
This equation can be used to calculate the total added-up probability for a known scale factor and known field modes for $\mb p=0$. Without this restriction, the resulting integral for the total probability would be extremely difficult to solve.  Next, we will apply this formula for the special cases of a universe dominated with radiation and matter.

\section{Radiation Era}\label{sec:III}
The radiation dominated era is described by a scale factor which scales with standard time as $a(t)\propto t^{1/2}$. In conformal time this corresponds to choosing the scale factor to be $a(\eta)^2=b^2\eta^2$, where $b$ is a positive constant controlling the expansion rate of the universe. The mode solutions for a universe dominated by radiation are known and are given in terms of parabolic cylinder functions $D_\alpha$  as \citep{Audretsch_Schafer:1978}
\begin{align}
\chi_p(\eta)=(2 mb)^{-1/4}e^{-\frac{\pi p^2}{8mb}}D_{-\frac{i p^2}{2mb}-\frac{1}{2}}((i+1)\sqrt{m b}\eta).  
\end{align}
In calculating the decay rate, we are only interested in the rest frame mode $\mb p=0$, so that the field mode is reduced to
\begin{align}
\chi_{p=0}(\eta)=(2 mb)^{-1/4}D_{-1/2}((i+1)\sqrt{m b}\eta).
\end{align}
With the change of variables $\sqrt{2mb}\eta=u$, the transition probability \eqref{Result1} takes the form
\begin{align}
\wt=\frac{\lambda^2}{32\pi m^2}I^t_{\rm{rad}}, 
\end{align}
where
\begin{align}\label{I_rad}
I^t_{\rm{rad}}=\int_0^{\sqrt{4mt}}u^2D_{-1/2}(e^{i\pi/4}u)D_{-1/2}(e^{-i\pi/4}u)du.
\end{align}
The $t$ in the upper limit of the integral refers to the standard coordinate time. To proceed, we write the parabolic cylinder functions in integral form using equation $9.245(2)$ from \cite{Gradstein}, so that
\begin{align}
I^t_{\rm{rad}}=\int_0^{\sqrt{4mt}}\int_0^\infty \frac{u^2}{\Gamma(1/2)}\frac{e^{\frac{u^2}{2}\sinh 2z}}{\sqrt{\coth z\sinh z}}dz du.
\end{align}
Making a change of variables $2w=\sinh 2z$ and after that $x=u^2w$, we have
\begin{align}\nonumber\label{I_radExact}
I^t_{\rm{rad}}&=\frac{1}{\sqrt{\pi}}\int_0^{\sqrt{4mt}}\int_0^\infty \frac{u^3e^{-x}}{\sqrt{4x^2+u^4}\sqrt{x}}dxdw\\
              &=\frac{1}{\sqrt{\pi}}\int_0^\infty \frac{e^{-x}}{\sqrt{x}}(-x+\sqrt{4m^2t^2+x^2})dx.
\end{align}
This integral can be integrated in exact form in terms of Bessel functions,

\begin{widetext}
\begin{align}\nonumber\label{eq:RadExact}
I_{\rm{rad}}^t=&m^2t^2\pi[\J{-1/4}{mt}^2-\sqrt{2}\J{-1/4}{mt}\J{1/4}{mt}+\J{1/4}{mt}^2+\J{5/4}{mt}^2-\J{3/4}{mt}\Y{3/4}{mt}]\\
&+\frac{mt\pi}{2\sqrt{2}}[\sqrt{2}\J{-3/4}{mt}\J{1/4}{mt}-\J{1/4}{mt}\J{3/4}{mt}+\J{-1/4}{mt}\Y{-3/4}{mt}]-\frac{\pi}{4}\J{1/4}{mt}^2-\frac{1}{2}.
\end{align}
\end{widetext}
Besides the exact form, we can also study the asymptotic behavior of the total probability. Therefore, we expand the integrand of Eq. \eqref{I_radExact} in asymptotic series in $t$. Taking only the leading terms in the asymptotic expansion, we are left with
\begin{align}\nonumber
I^t_{\rm{rad}}&\sim\frac{1}{\sqrt{\pi}}\int_0^\infty\frac{e^{-x}}{\sqrt{x}}(-x+2mt)dx\\
              &=2\Big(mt-\frac{1}{4}\Big).
\end{align}
Hence, the leading two terms in the asymptotic expansion of the total transition probability are given by
\begin{align}
\wt\sim\frac{\lambda^2}{16\pi m}\Big(t-\frac{1}{4m}\Big).
\end{align}
The decay rate in flat spacetime is obtained by dividing the infinite transition probability by the time $t$. In curved space this procedure is more complicated because the probability contains an additive term. We will proceed as in Ref. \citep{Audretsch_Spangehl:1985}, where the additive term is divided by a finite time $t_{\rm{grav}}$ representing the time of gravitational influence. It can be defined as $t_{\rm{grav}}:=t_f-t_i$, where $t_i$ denotes the time when the gravitational field begins its influence and $t_f$ its end. The mean decay rate for a radiation dominated universe is then 
\begin{align}
\Gamma_{\rm{rad}}\sim\frac{\lambda^2}{16\pi m}\Big(1-\frac{1}{4mt_{\rm{grav}}} \Big).
\end{align}

\section{Matter Era}\label{sec:IV}
The scale factor which scales with the standard time as $a(t)\propto t^{2/3}$ describes a universe dominated by ordinary matter. The corresponding scale factor in conformal time can be written as $a(\eta)^2=b^2\eta^4$, with $b$ being a positive constant. For this choice of the scale factor, the mode equation \eqref{KG_chi} does not possess a solution in terms of known functions. The general modes for a matter dominated universe are, therefore, not known. In calculating the decay rate using the added-up formalism however, we are ultimately using only the rest frame modes. Although the generalized field modes, for example for the radiation dominated universe, are known, in the end $\mb p=0$ is set. 
With this in mind, we solve the mode equation \eqref{KG_chi} for the special case of $p=0$, obtaining the mode required for calculation of the decay rate.
Choosing $a(\eta)^2=b^2\eta^4$, the Eq. \eqref{KG_chi} for the rest frame field modes reads as
\begin{align}\label{Matter_diffis}
\chi''_{ p=0}(\eta)+m^2b^2\eta^4\chi_{ p=0}(\eta)=0.
\end{align}
The general solution for Eq. \eqref{Matter_diffis} is given in terms of Bessel functions as
\begin{align}
\chi_{ p=0}(\eta)=&c_1J_{-1/6}\left(\frac{1}{3} b m \eta ^3\right)+c_2J_{1/6}\left(\frac{1}{3} b m \eta ^3\right),
\end{align}
where $c_1$ and $c_2$ are constants and $J_\alpha$ denotes the Bessel function of the first kind. The positive mode can be recognized by examining the asymptotic behavior of the mode and the normalized positive mode is found to be
\begin{align}\label{MatterMode}
\chi_{ p=0}(\eta)= e^{5 i\pi/3}\frac{\sqrt{\pi \eta}}{12}H^{(2)}_{1/6}\Big( \frac{b m\eta^3}{3}\Big),
\end{align}
where $H^{(2)}_\alpha$ denotes the Hankel function of the second kind. Making a change of variables $u=bm\eta^3/3$, the transition probability \eqref{Result1} takes the form
\begin{align}
\wt=\frac{\lambda^2}{32\pi m^2}I_{\rm{mat}}^t,
\end{align}
where
\begin{align}\label{I_mat}
I_{\rm{mat}}^t=\pi\int_0^{mt}u H^{(2)}_{1/6}(u)H^{(1)}_{1/6}(u)du.
\end{align}
The variable $t$ is again the standard coordinate time and $H^{(1)}_\alpha$ denotes the Hankel function of the first kind. The integral \eqref{I_mat} can be cast in the form
\begin{align}
I_{\rm{mat}}^t=\pi\int_0^{mt}u [J_{1/6}(u)^2+Y_{1/6}(u)^2]du
\end{align}
and using equation $5.54(2)$ from \cite{Gradstein}, this can be integrated yielding
\begin{align}\nonumber\label{I_mat_b}
I_{\rm{mat}}^t=&\frac{\pi m^2t^2}{2}[J_{1/6}(mt)^2-J_{-5/6}(mt)J_{7/6}(mt)\\
&+Y_{1/6}(mt)^2-Y_{-5/6}(mt)Y_{7/6}(mt)]-\frac{1}{\sqrt{3}}.
\end{align}
Examining the asymptotic form of this, we find that the first terms in the asymptotic series are
\begin{align}
I_{\rm{mat}}^t\sim 2m\Big(t-\frac{1}{2\sqrt{3}m} \Big).
\end{align}
Proceeding as in the radiation case, we find the mean decay rate in the matter dominated era to be
\begin{align}
\Gamma_{\rm{mat}}\sim\frac{\lambda^2}{16\pi m}\Big(1-\frac{1}{2\sqrt{3}mt_{\rm{grav}}} \Big).
\end{align}

\section{Discussion}\label{sec:V}
In the preceding sections, we have derived the mean decay rate for a massive particle decaying into two massless particles in universes dominated by radiation and matter. In our previous article Ref. \citep{Lankinen_Vilja:2017b} we have done the same for a universe filled with stiff matter. Collecting these results together, the leading two asymptotic terms in these decay rates are
\begin{align}\label{Gamma1}
\Gamma_{\rm{mat}}&\sim\frac{\lambda^2}{16\pi m}\Big(1-\frac{1}{2\sqrt{3}mt_{\rm{grav}}} \Big),\\\label{Gamma2}
\Gamma_{\rm{rad}}&\sim\frac{\lambda^2}{16\pi m}\Big(1-\frac{1}{4mt_{\rm{grav}}} \Big), \\\label{Gamma3}
\Gamma_{\rm{stiff}}&\sim\frac{\lambda^2}{16\pi m}\Big(1-\frac{1}{3\sqrt{3}mt_{\rm{grav}}} \Big).
\end{align}
These can be compared with the Minkowskian decay rate
\begin{align}
\Gamma_{\rm{Mink}}=\frac{\lambda^2}{16\pi m},
\end{align}
which is obtained by the same added-up formalism.
All these decay rates share the same feature of having a Minkowskian part corrected by a finite gravitational part proportional to the inverse of mass. Moreover, the sign of the correction is negative in all and proportional to the inverse of mass implying that the decay rate is smaller and lifetime of the particles longer when compared to flat Minkowskian space. The asymptotic formulas \eqref{Gamma1}, \eqref{Gamma2} and \eqref{Gamma3} only valid when positive, which means that $t_{\rm{grav}}$ is restricted to be greater than a small constant times the inverse of mass. Since $t_{\rm{grav}}$ is usually much longer than the inverse of mass, these correction terms are in practice very small. Considering a practical cosmological setting, it therefore seems that the correction terms is not really significant when compared to the Minkowskian term. The relative correction terms might not be neglected altogether, however, since for $t\sim m$ the full equations should be used instead of the asymptotic expansions.

\begin{figure}[H]
\centering
\includegraphics[width=1.0\columnwidth]{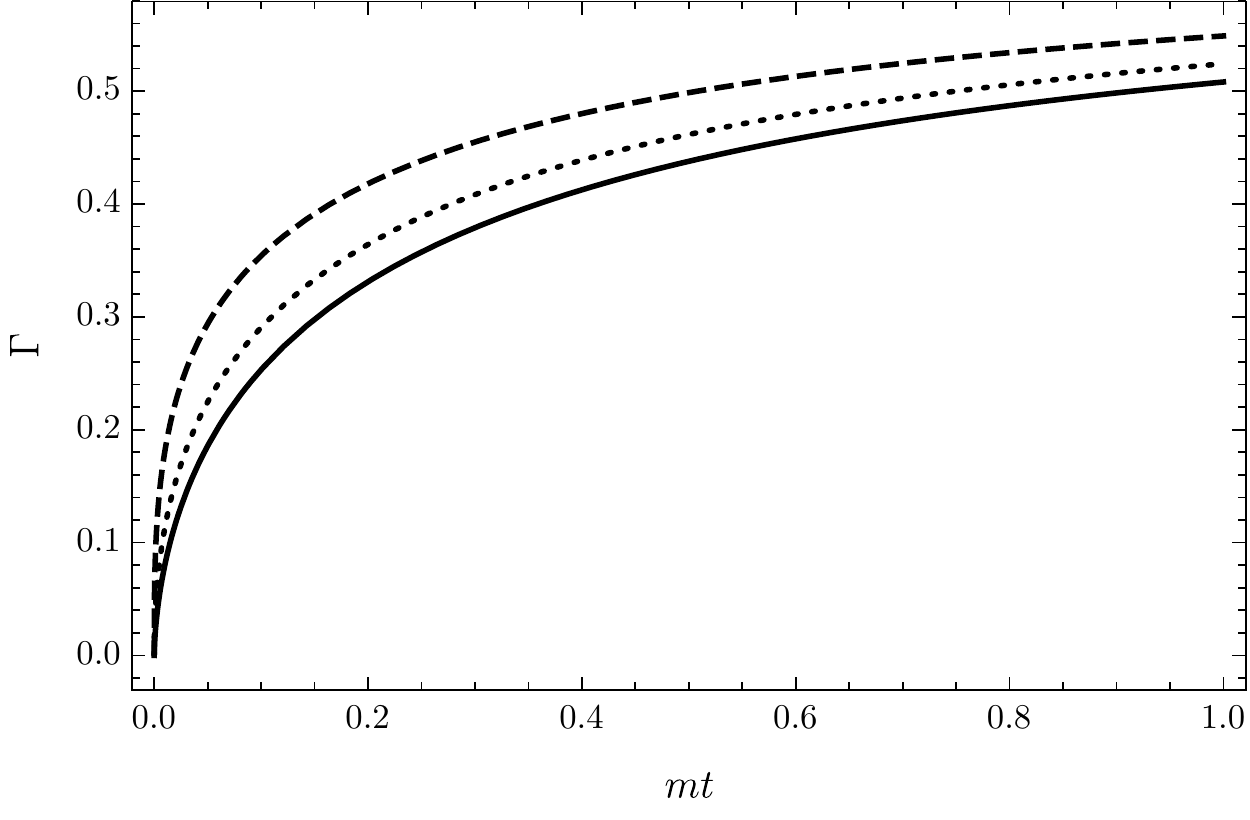}
\caption{Exact decay rate $\Gamma$ as a function of $mt$. Dashed line corresponds to stiff matter, dotted to radiation and solid line to matter, respectfully.}\label{fig:2}
\end{figure}

A peculiar feature is found when examining the curved space decay rates. They appear to be organized in a way that the correction is largest for matter dominated universe and smallest for stiff matter dominated universe with radiation lying between these two. The differences between these corrections are, however, very small but still worth noticing. As the time of the gravitational influence approaches infinity, these decay rates approach that obtained in flat space. Although the asymptotic forms show this ordering for these three matter contents, it actually holds also when probing small time scales (Fig. \ref{fig:2}), when the full exact decay rate of \eqref{Gamma1}, \eqref{Gamma2} is used along with the exact decay rate of \eqref{Gamma3} obtained from Eq. (21) of Ref. \citep{Lankinen_Vilja:2017b}.

We may speculate the cause of this organization. It might be related to the expansion rate of the universe in question. Indeed, we can inspect the Hubble parameter $H$, which is given by
$H=\alpha/t$, with $\alpha=1/3$ corresponding to stiff matter, $\alpha=1/2$ to radiation and $\alpha=2/3$ to ordinary matter. From these we can infer that for a matter dominated era, for which the relative correction term is largest, also the relative expansion rate is greatest. This reflects the idea that the faster the Universe is expanding the smaller is the decay rate. 
It cannot be said that this is truly the cause for the differences in the decay rates and further investigations into this feature are definitely needed.

The results for the decay rates warrant a lot of discussion; like why the decay rate has an explicit Minkowskian part in the $t_{\rm{grav}}^{-1}$ expansion. Also there are some subtle issues regarding the fact that the metrics used in these calculations do not have a well defined Minkowskian limit, or the fact that the asymptotic forms of the decay rates are not always positive. We have addressed these issues in great detail in our previous article Ref. \citep{Lankinen_Vilja:2017b}, which are applicable also to the cases of matter and radiation dominated universes.

All things considered, the presence of a gravitational field alters the decay rates usually calculated in flat Minkowskian spacetime. In case of radiation, matter and stiff matter dominated universe, there is an additive correction proportional to the inverse of mass. Surprisingly, this correction term is insignificant at large time scales making the Minkowskian approximation sound even in curved spacetime for this process.

\begin{acknowledgments}
J.L would like to acknowledge the financial support from the University of Turku Graduate School (UTUGS).
\end{acknowledgments}
\bibliography{references}

\end{document}